# Weighted Jump in Random Walk Graph Sampling

Xiao Qi

x.qi274@gmail.com

## Abstract

Random walk based sampling methods have been widely used in graph sampling in recent years, while it has bias towards higher degree nodes in the sample. To overcome this deficiency, classical methods such as GMD modify the topology of target graphs so that the long-term behavior of Markov chain can achieve uniform distribution. This modification, however, reduces the conductance of graphs, thus makes the sampler stay in the same node for long time, resulting in undersampling. To address this issue, we propose a new way of modifying target graph, thus propose Weighted Jump Random Walk (WJRW) with parameter C to improve the performance. We prove that WJRW can unify Simple Random Walk and uniform distribution through C, and we also conduct extensive experiments on real-world dataset. The experimental results show WJRW can promote the accuracy significantly under the same budget. We also investigate the effect of the parameter C, and give the suggested range for a better usage in application.

## 1. Introduction

In recent years, graphs have become increasingly used as the basic representation of data that appears in various domains. Examples include road networks (Duan and Lu, 2014; Xie and Levinson, 2007), communication networks (Gupta et al., 2016; Shimbel, 1953), online social and professional networks (Ahn et al., 2007; Bos et al., 2018), citation networks (McLaren and Bruner, 2022; Portenoy et al., 2017), collaborative networks (Newman, 2001), biological networks (Charitou et al., 2016; Zhang and Itan, 2019), etc. However, handling large-scale graphs and efficiently solving graph mining problems are not trivial tasks. To meet this challenge, sampling methods have been developed. This is a technique that investigate small percentage of online social network (OSN) users or relations to estimate the characteristics of OSN.

In graph sampling techniques, Simple Random Walk (SRW) based methods made up of large proportion in sociocentric methods. In RW-based sampling methods, the most commonly used methods are Simple Random Walk (SRW), and Maximum Degree (MD).

SRW is the most basic model on graphs. SRW are used to collect data and the estimator is constructed based on importance sampling framework, reweighting the sample by the inverse of inclusion probabilities. In this way, we can perform an unbiased estimation for the function of interest. However, the inclusion probabilities of sample, i.e., the value of stationary distribution as a function at specific node (proportional to node degree), varies in a wide range, due to the real-world networks usually follow power-law (Adamic et al., 2008). The similarity between stationary distribution of SRW and the uniform distribution determines the effectiveness of random walk (Li et al., 2015), so one main issue in proposing variants of RW-based sampling methods is to increase the similarity between the stationary distribution and uniform distribution.

One classical modification called Maximum Degree (MD) (Bar-Yossef et al., 2000; Bar-Yossef and Gurevich, 2008) is adding self-loop in low-degree nodes, and the numbers of nodes depend on the difference between the maximum degree of the graph $d_{max}$ and the node degree $d_i$. After adding self-loops on nodes, all nodes in the graph have the same degree $d_{max}$, thus performing SRW on the modified graph can get a uniform stationary distribution. However, MD may lead to large proportion of repeated sample due to self-loops, thus result in low accuracy in estimation. Li et al. (2015) improved MD by adding a parameter $C$ into MD to adjust the number of self-loops by $C - d_i$ at node $i$. When $C = d_{max}$, the new algorithm is the same as the traditional MD. When $C < d_{max}$, GMD can significantly alleviate the repeated sample problems, while the stationary distribution is not uniform. Although GMD alleviate the repeated sample, the existence of self-loops always leads to repeated sample problems. In a word, GDM suffers whether large deviation problems or repeated sample problems.

**Our contributions.** To address the limitations in previous sampling methods, we propose Weighted Jump Random Walk (WJRW) in this paper. Our new method can balance the tradeoff between the large deviation problem of SRW and the cost of uniform distribution, treating SRW and uniform distribution as two special cases in the WJRW framework. We prove that WJRW does alleviate repeated samples problems theoretically and numerically, and can also plays as a framework unifying Simple Random Walk (SRW) and uniform distribution. That is similar with the GMD framework, but can do a better job than GMD under the same parameter chosen situations. The experimental results also show our method outperform other algorithms in both large and medium-scale datasets.

**Organization.** The rest of the paper is organized as follows: Section 2 is the notation that will used in this paper, and the preliminaries of RW-based sampling strategies. Section 3 presents our new methods in theory and the benefits, and Section 4 is the analysis of numerical simulation, discussing why WJRW outperforms other methods. Section 5 is the conclusion and future research.

## 2. Notation and preliminaries

In Section 2, we introduce notations used in this paper, and the basic facts of RW-based sampling methods, including Simple Random Walk (SRW), Random Walk with Escaping (RWE), Maximum Degree (MD), and Generalized Maximum Degree (GMD), and the estimation framework when they are used in graph sampling.

### 2.1 Basic Notations

If we have an undirected and connected graph that can be denoted by $G\ (V, E)$, where $V$ is the node set and $E$ is the edge set. As commonly used statistics of graphs, the number of node and edges of $G$ are denoted by $|V|$ and $|E|$, respectively. For convenience, we also define neighbor set of a node $v$ as $N(v)$, and the degree of $v$, i.e., the number of neighbors of $v$, is denoted as $d_v$.

Let $f: V \mapsto \mathbb{R}$ be a characteristic function of interest that maps a node in the graph to a real number. The goal of measuring the characteristic function of $G$ is to estimate

$$\theta \triangleq \frac{1}{n} \sum_{v \in V} f(v)$$

$\theta$ is an aggregated nodal characteristic of the graph. For example, for a specific degree $k$, we set $f(v) = 1$ if $d_v = k$, otherwise $f(v) = 0$. Then, $\theta$ is the fraction of nodes with degree $k$ in the whole graph. If we calculate all possible $k$ in the graph, then we can get the degree distribution of the graph.

### 2.2 Simple Random Walk on graphs

From the diffusion aspect, random walk (RW) on graph $G$ is a procedure that starts from one node, say $v_0 \in V$, and randomly chooses a neighbor of the current node as the next node. Moving to the nodes step by step makes up of a complete random walk on a graph. Random walk can be comprehended in an abstract way; it can be viewed as a discrete- time Markov chain in finite state space. The state space contains all nodes in the node set $V$, and the edges, which link different states of the Markov chain, represents different probability in different selection mechanisms.

Here we introduce Simple Random Walk (SRW), which is the foundation of random walk algorithms. The selection mechanism is uniformly choosing a neighbor of current node $v$, so in SRW, the probability of moving from $v$ to $u$ is

$$P = \begin{cases} \frac{1}{d_v}, & if\ u \in N(v) \\ 0, & otherwise \end{cases}$$

Note that the graph is not equivalent to the Markov chain diagram. For an undirected and connected graph $G$, walking from $u$ to $v$ depends on $d_u$ while walking from $v$ to $u$ depends on $d_v$. The adjacent matrix of a graph $G$ is denoted by $A$.

$$A = \begin{pmatrix} a_{11} & \cdots & a_{1N} \\ \vdots & \ddots & \vdots \\ a_{N1} & \cdots & a_{NN} \end{pmatrix}$$

thus, the walk matrix, also the transition matrix of SRW is

$$P = \begin{pmatrix} \frac{a_{11}}{\sum a_{1\cdot}} & \cdots & \frac{a_{1N}}{\sum a_{1\cdot}} \\ \vdots & \ddots & \vdots \\ \frac{a_{N1}}{\sum a_{N\cdot}} & \cdots & \frac{a_{NN}}{\sum a_{N\cdot}} \end{pmatrix}$$

where $\sum a_{k\cdot}$ represents the sum of all elements in $k$th row in $A$, and $p_{vu}$ in $P$ represents the probability that random walker can walk from node $v$ to node $u$ in one step. Therefore, we can make use Markov chain theory to get the stationary distribution of the process. Given an initial distribution $\pi_0$, which represents the initial proportion of visited nodes, we use the detailed balance condition

$$\pi_v P_{uv} = \pi_v P_{vu}$$

since $\pi = (\pi_1, \dots, \pi_n)$ is a probability density function, it satisfies the condition $\sum_{i=1}^{N} \pi_i = 1$, to calculate the stationary distribution of the Markov chain. Finally, we can make use of the stationary distribution of the Markov chain

$$\pi_v = \frac{d_v}{\sum_{v \in V} d_v} = \frac{d_v}{2|E|}.$$

to derive unbiased estimators of the target network properties.

One limitation of SRW is that the random walker cannot reach equilibrium on disconnected graphs, so the Random Walk with Escaping (RWE) (Bianchini et al., 2005; Jin et al., 2011; Leskovec and Faloutsos, 2006) is proposed to overcome the limitation of SRW. The jumping part is uniformly distributed so the RWE is a mixture of random walk and uniform node sampling. Setting a parameter $\alpha$ as the weight of edge linking node $v$ and a virtual node, which would trigger the jumping mode in node $v$, that $\frac{\alpha}{d_v+\alpha}$ is the probability of jumping while $\frac{d_v}{d_v+\alpha}$ is the probability of walking. Therefore, the stationary distribution of RWE is

$$\pi_v = \frac{d_v + \alpha}{\sum(d_v + \alpha)}.$$

From the stationary distribution we can also see that adding an $\alpha$ can alleviate differences between high-degree nodes and low-degree nodes. The larger $\alpha$ is, the closer to uniform the stationary distribution of RWE is. It is interesting that whether the inclusion probability of $v$ collected by hybrid random walk is larger than that of simple random walk depends on the degree of $v$ and the average degree of the whole graph, which is defined as

$$<d>:= \frac{\sum_{v\in V} d_v}{|N|}$$

If $d_v$ is larger than $<d>$, the inclusion probability of $v$ would decrease in hybrid random walk, otherwise, the hybrid random walk would elevate the inclusion probability of a low-degree node compared with simple random walk. The increase or the decrease of the inclusion probability also depends on the value of alpha. The change of inclusion probability increases as the increasing value of $\alpha$.

An intuition is that with a large $\alpha$, jumping mode participate in a large extent, i.e., the proportion of uniform node sampling is large, while walking mode participates in a small proportion.

**2.3 Maximum Degree (MD) and Generalized Maximum Degree (GMD)**

MD (Bar-Yossef et al., 2000; Bar-Yossef and Gurevich, 2008) runs a maximum degree random walk to gather nodes, adding $C - d_v$ self-loops on node $v$, so that every node in the graph can get a node degree $C$, which is the maximum degree of the graph. As SRW's stationary distribution is proportional to the nodes' degrees, the stationary distribution of maximum degree random walk is $\pi_v = \frac{C}{\sum C} = \frac{1}{N}$, i.e., uniform distribution. Therefore, we can get a uniform sample from MD.

However, MD has two main limitations. First, we need to know the whole graph so that we can confirm the maximum degree $C$. Second, adding self-loops is too expensive to conduct in real-world application. Therefore, Li et al. (2015) proposed a generalized version of MD with adding nodes' degrees to $\max\{C, d_v\}$, instead of the maximum degree $d_{max}$ in MD. By setting an appropriate parameter, GMD can not only alleviate the large deviation problem of RW, but it is also expected to produce a smaller number of repeated samples compared to MD. It also gives a link between SRW and MD, while they are regarded as two different cases in the past. No matter how GMD improved MD, the main idea is to modify the topology of the graphs by adding self-loops on them. Although self-loops can increase the degree of low-degree nodes thus increase the inclusion probability via SRW, the essence of the increased probability is forcing the random walker staying in the same node, which is a disguised "being stuck". That is not good for the diffusion on the graphs, so the sampler would select a bad sample thus the estimates would not be good based on a poor sample. We would use an example to illustrate how self-loops lower the conductance of a graph in the Section 3.

**2.4 Unbiased Estimator**

Now that we have a sample collected by SRW, we need an estimator with good mathematical properties. Therefore, we can derive that if we employ SRW as the sampling method to collect nodes, the probability of being selected of a node is proportional to its degree. So, in the estimation framework, researchers usually use Horvitz–Thompson estimator (Horvitz and Thompson, 1952). When random walker reaches the steady state, a node $v$ is sampled with probability $\pi(v)$, where $\pi(v)$ differs according to the stationary distributions generated by neighbor selection mechanism. If we denote the set of sampled nodes as $S$, an asymptotically unbiased estimator of $\theta$ is given by

$$\hat{\theta} = \frac{\sum_{s \in S} \frac{f(s)}{\pi(s)}}{\sum_{s \in S} \frac{1}{\pi(s)}}$$

The theory backing up the unbiasedness of estimator is the Law of Large Numbers (LNN) of Markov chains.

**Lemma 1**: (Law of Large Number). Let $S$ be a sample path obtained by a Markov chain defined on a state space $V$ with stationary distribution $\pi$. For any function $f, g: U \mapsto \mathbb{R}$, and let $F_S(f) \triangleq \sum_{v \in V} \pi_v f(v)$. It holds that

$$\lim_{|S| \to \infty} \frac{1}{|S|} F_S(f) = E_\pi[f]$$

and

$$\lim_{|S| \to \infty} \frac{F_S(f)}{F_S(g)} = \frac{E_\pi[f]}{E_\pi[g]}$$

almost sure. Now that we can make use LNN, the estimated value $\hat{\theta} = \frac{\sum(f(s)/\pi(s))}{\sum(1/\pi(s))}$ converges to $\theta$. So, estimator (1) is an asymptotically unbiased estimator.

# 3. Random Walk with Weighted Jump

In this section, we propose a novel weighted jump random walk (WJRW) algorithm for unbiased graph sampling. we show that the WJRW algorithm can balance the repeated sample problem of GMD and large deviation problem of SRW. Below, we first describe the basic idea of our algorithm. Then, we present the WJRW algorithm and the theoretical analysis as well. Finally, we derive an unbiased estimator given the sample from WJRW.

## 3.1 Main idea of Weighted Jump Random Walk

To overcome the deficiency of proposed methods, we proposed a new method Weighted Jumps, whose idea is originated from the RWE and GMD. Similar with GMD, we also introduce a controlled parameter $C$ into our algorithm. Instead of adding self-loops in the low-degree nodes, we set a virtual node and add $\max\{C, d_v\} - d_v$ edges from $v$ to the virtual node and run a simple random walk on the graphs. The virtual node is a status that will trigger jumping mode i.e., uniformly select a node connected with it as the next node. As the number of edges linking node $v$ and virtual node depends on the difference between $d_v$ and $C$, the probability of jumping mode is $(\max\{C, d_v\} - d_v)/\max\{C, d_v\}$, and the probability of walking mode is $d_v/\max\{C, d_v\}$. This is different from the setting of jumping mode in aforementioned RWE, as RWE assume every node in the graph is connected with the virtual node, so that every node has a prefixed and same probability $\frac{\alpha}{\alpha+d_v}$ to walk to the virtual node, triggering jumping mode. Therefore, some high-degree nodes also add the probability of being reached by jumping mode, and the probability can only be alleviated by reassigning probabilities by a larger denominator.

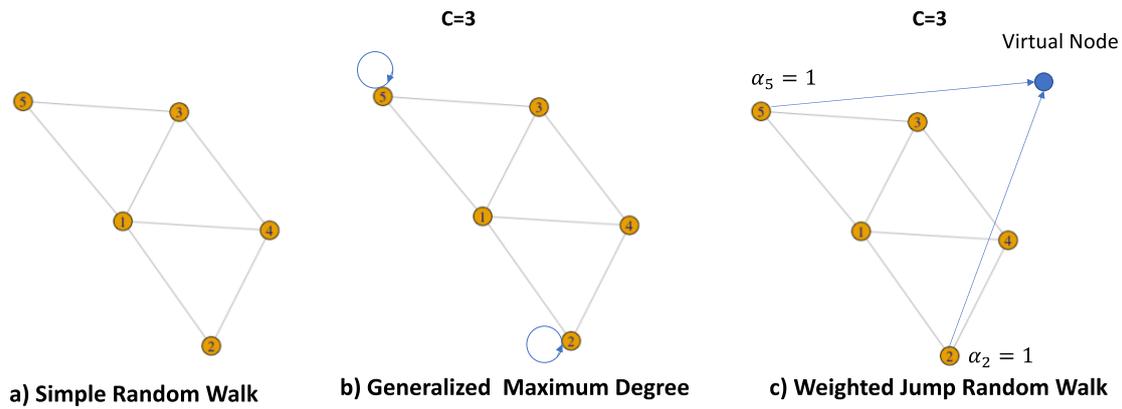

*Figure 1 A simple example*

We would use a simple example to illustrate the deficiency of the existence of self-loop. Fig. 1 is a simple graph with 5 nodes. For SRW, the transition probability is walk matrix.

$$P^{SRW} = \begin{pmatrix} 0 & \frac{1}{4} & \frac{1}{4} & \frac{1}{4} & \frac{1}{4} \\ \frac{1}{2} & 0 & 0 & \frac{1}{2} & 0 \\ \frac{1}{3} & 0 & 0 & \frac{1}{3} & \frac{1}{3} \\ \frac{1}{3} & \frac{1}{3} & \frac{1}{3} & 0 & 0 \\ \frac{1}{2} & 0 & \frac{1}{2} & 0 & 0 \end{pmatrix}$$

For GMD, assume $C=3$, the one-step transition matrix is

$$P^{GMD} = \begin{pmatrix} 0 & \frac{1}{4} & \frac{1}{4} & \frac{1}{4} & \frac{1}{4} \\ \frac{1}{3} & \frac{1}{3} & 0 & \frac{1}{3} & 0 \\ \frac{1}{3} & 0 & 0 & \frac{1}{3} & \frac{1}{3} \\ \frac{1}{3} & \frac{1}{3} & \frac{1}{3} & 0 & 0 \\ \frac{1}{3} & 0 & \frac{1}{3} & 0 & \frac{1}{3} \end{pmatrix}$$

And for WJRW, the one-step transition matrix is

$$P^{WJRW} = \begin{pmatrix} 0 & \frac{1}{4} & \frac{1}{4} & \frac{1}{4} & \frac{1}{4} \\ \frac{1}{3} & \frac{1}{6} & 0 & \frac{1}{3} & \frac{1}{6} \\ \frac{1}{3} & 0 & 0 & \frac{1}{3} & \frac{1}{3} \\ \frac{1}{3} & \frac{1}{3} & \frac{1}{3} & 0 & 0 \\ \frac{1}{3} & \frac{1}{6} & \frac{1}{3} & 0 & \frac{1}{6} \end{pmatrix}$$

Now we calculate the eigenvalues of walking matrix, the second-largest eigenvalue of $P^{WJRW}$, $P^{SRW}$ and $P^{GMD}$, denoted by $\mu^{WJRW}$, $\mu^{SRW}$ and $\mu^{GMD}$. $\mu^{WJRW} = \frac{\sqrt{5}-1}{6}$, $\mu^{SRW} = \frac{\sqrt{7}-1}{6}$, and $\mu^{GMD} = \frac{\sqrt{2}}{3}$, and we have $\mu^{WJRW} < \mu^{SRW} < \mu^{GMD}$. This quantitative relation between three second-largest eigenvalues implies the WJRW converges faster than SRW, and SRW converges faster than GMD. MD/ GMD was initially proposed to overcome large deviation problems in SRW; however, the self-loops result in both repeated sample problems and slow convergence problems. If we regard "graph sampling + accurate estimation" as the final goal, MD/GMD did not achieve the goal very well.

### 3.2 Analysis of stationary distribution

The stationary distribution plays an important role in RW-based sampling method, as stationary distribution would be utilized as the sampling probability, in which the inclusion probabilities of nodes and/ or edges is captured from.

To get the stationary distribution, in this part, we would start from the transition matrix of WJ, and prove that the uniqueness and existence of stationary distribution of WJ, thus make use of detailed balance to derive the stationary distribution. For an undirected and connected graph $G$, once some of nodes with lower degree than $C$ are connected with the virtual node, we denote

$$U = \{v|\ v\ is\ connected\ with\ virtual\ node\} = \{v|d_v < C\}$$

Thus, the number of nodes linking virtual node is $|U|$. Nodes in $U$ have access to all nodes connected with the virtual node with probability $1/|U|$, while nodes with higher degree than $C$ can only access to their own neighbors existing in the graph.

The transition probability from $v$ to $u$ can be divided into 2 parts: jumping mode and walking mode.

1) The jumping mode: as we mentioned above, node $v$ has probability $\frac{\max\{C,d_v\}-d_v}{\max\{C,d_v\}}$ to trigger the jump mode, and select a node in $U$ with probability $\frac{1}{|U|}$, so if $v$ moves to $u$ through jumping mode, the probability is $\frac{\max\{C,d_v\}-d_v}{\max\{C,d_v\}}\frac{1}{|U|}\mathbb{I}\{u \in U\}$, where $\mathbb{I}\{u \in U\}$ is an indicator to judge whether node $u$ is connected with the virtual node, i.e., whether node $u$ can be reached by jumping mode.

2) The walking mode: if $v$ chooses the walking mode, and $u$ is a neighbor of $v$, then the transition probability is $\frac{d_v}{\max\{C,d_v\}}\frac{1}{d_v} = \frac{1}{\max\{C,d_v\}}$. Also, not all the node is neighbor of $v$; without edges between $u$ and $v$, even $v$ chooses walking mode, the transition probability from $v$ to $u$ by walking equals to zero. So, we need to add an indicator to judge whether $u$ is connected with $v$. in the adjacent matrix $A$, $a_{vu}$ represents whether there is an edge between $u$ and $v$. Thus, the transition matrix by walking is $\frac{a_{vu}}{\max\{C,d_v\}}$.

Combining two kinds of probabilities, we have the transition probability from $v$ to $u$ is

$$P_{vu} = \frac{\max\{C,d_v\}-d_v}{\max\{C,d_v\}}\frac{1}{|U|}\mathbb{I}\{u \in U\} + \frac{a_{vu}}{\max\{C,d_v\}}$$

where $a_{vu}$ is the element in the $v$-th row, $u$-th column in the adjacent matrix $A$, $U$ is the node set containing nodes with degrees lower than $C$, and $\mathbb{I}$ is an indicator function. Now we need to derive the stationary distribution of Weighted Jump Random Walk.

**Theorem 1:** If graph $G = (V,E)$ is connected and undirected, Weighted Jump Random Walk has a unique stationary distribution on $G$.

***Proof:*** For any node $v$ and $u \in N(v)$, the transition probability $P_{vu}$ is larger than 0, as the walking mode is always positive. Thus, for any two nodes $u$ and $v$ in $G$, they are reachable from each other in finite steps for Weighted Jump since $G$ is an undirected and connected graph. therefore, we can conclude that the Markov chain constructed by Weighted Jump Random Walk is irreducible. According to Markov chain theory, any irreducible Markov chain on an undirected and connected graph has a unique distribution, WJRW has a unique stationary distribution.

**Theorem 2:** The stationary distribution of WJRW on $G$ is

$$\pi_v^{WJRW} = \frac{d_v'}{\sum_{v \in V} d_v'}$$

where $d_v' = d_v + \mathbb{I}\{d_v < C\}\frac{\sum_{u \in V}\alpha_u}{\sum_{u \in V}\mathbb{I}\{d_u < C\}}$, and $\alpha_u = \max\{C,d_u\} - d_u$.

***Proof:*** The virtual node plays as a transfer station when low-degree nodes in the graph chooses to jump, instead of a real node in the graph. So WJRW can be recognized as a weighted random walk that still follows the selection mechanism of SRW, but on a modified graph. Apart from the original edges in the graph, every low-degree node has an edge, whose weight equals to $\alpha_v$, connecting to the virtual node, and reach the nodes in $U$ with equal probability $1/|U|$. therefore, as long as a node is connected to the virtual node, it has an edge that the weight $\frac{\sum_{u \in V} \alpha_u}{\sum_{u \in V} \mathbb{I}\{d_u < C\}}$, while nodes with higher degree do not have this mode. Therefore, we add an indicator to decide whether add the probability of jumping mode or not.

As nodes in the graph also have walking mode to choose, the weight of edge in the original graph is equivalent to $d_v$. Sum the jumping and walking mode, edges connected node $v$ in WJRW equals to $d_v + \mathbb{I}\{d_v < C\}\frac{\sum_{u \in V} \alpha_u}{\sum_{u \in V} \mathbb{I}\{d_u < C\}}$, where $\alpha_u = \max\{C, d_u\} - d_u$.

From theory of Weighted Random Walk model, if we know the weights of edges connecting to the node $i$, which is denoted by $w_i$, the stationary distribution of the graph is

$$\pi_i = \frac{w_i}{\sum w_i}$$

In this way, we can write the stationary distribution as $\pi_v^{WJRW} = \frac{d_v + \mathbb{I}\{d_v < C\}\frac{\sum_{u \in V} \alpha_u}{\sum_{u \in V} \mathbb{I}\{d_u < C\}}}{\sum\left(d_v + \mathbb{I}\{d_v < C\}\frac{\sum_{u \in V} \alpha_u}{\sum_{u \in V} \mathbb{I}\{d_u < C\}}\right)}$, and $\alpha_u = \max\{C, d_u\} - d_u$. For simplicity, we denote $d'_v = d_v + \mathbb{I}\{d_v < C\}\frac{\sum_{u \in V} \alpha_u}{\sum_{u \in V} \mathbb{I}\{d_u < C\}}$, thus $\pi_v^{WJRW} = \frac{d'_v}{\sum_{v \in V} d'_v}$. ∎

Different from GMD, WJRW adds edges to a virtual node, instead of adding them into low-degree nodes themselves. Thus, WJRW gives a smaller expected traversing the self-loops in the next step, overcoming the disguised "stuck" and alleviate the repeated sample problem caused by self-loop in MD and GMD. Following Theorem 3 shows that WJRW has good mathematical properties –alleviate the repeated sample problem. To measure the repeated sample problem, we investigate the probability of reaching $v$ in the next step given the current node is $v$, which is denoted by $r(v)$. For GMD and WJRW, the probabilities of staying in node $v$ in two consecutive states are denoted by $r_{GMD}(v)$ and $r_{WJRW}(v)$, respectively. The expectations, are denoted by the $E_{\pi_{GMD}}[r_{GMD}(v)]$ and $E_{\pi_{WJRW}}[r_{WJRW}(v)]$.

**Theorem 3:** Given $C < d_{max}$, $E_{\pi_{GMD}}[r_{GMD}(v)] \geq E_{\pi_{WJRW}}[r_{WJRW}(v)]$.

*Proof:* By the definition of expectation,

$$E_{\pi_{WJRW}}[r_{WJRW}(v)]$$
$$= \sum_{v \in V} \frac{\max\{d_v, C\}}{\sum_{u \in V} \max\{d_v, C\}} \frac{\max\{d_v, C\} - d_v}{\max\{d_v, C\}} \frac{1}{|S|}$$
$$= \frac{1}{|S|} E_{\pi_{GMD}}[r_{GMD}(v)]$$

$$\leq E_{\pi_{GMD}}[r_{GMD}(v)]$$

$$\leq E_{\pi_{MD}}[r_{MD}(v)]$$

Theorem 3 implies that for low-degree nodes, Weighted Jump Random Walk has a smaller probability staying in the same node compared with the GMD and MD. Therefore, WJRW can alleviate the repeated samples than the GMD and MD. In the experiments, we shall show that the WJRW degreases the number of repeated samples compared to GMD and MD.

### 3.3 Unbiased estimator in graph sampling

After obtaining the sample by Algorithm 1, we can construct an estimator for the function of interest $\theta$ by

$$\hat{\theta} = \frac{\sum_{s \in S} \frac{f(s)}{\pi(s)}}{\sum_{s \in S} \frac{1}{\pi(s)}}$$

where $\pi(s)$ is the probability value of a stationary distribution of node $s$, and $f(s)$ is the function value at node $s$.

## 4. Experimental Evaluation

In this section, we conduct various numerical experiments on both real-world datasets. Our experiments include 3 parts. The first part is drawing samples from large-scale real datasets and compare the estimation accuracies of 4 different sampling algorithms. The second part is to investigate the rate of convergence of different RW-based algorithms using small datasets. In the last part, we would change the value of $C$ and see how $C$ affects the estimation accuracy. All the methods are implemented in R and tested on a PC with 1.6 GHz Dual-Core Intel Core i5.

### 4.1 Experimental Setup

In our simulations, we use 4 publicly accessible real-world datasets: YouTube, DBLP, Wiki, and Slashdot. (1) YouTube is a network dataset representing the friendship between users in YouTube(Yang and Leskovec, 2012). (2) DBLP is a coauthor network where two authors are connected if they publish paper together (Yang and Leskovec, 2012). (3) Wiki is a Wikipedia page-page network focusing on crocodile in Wikipedia (Rozemberczki et al., 2021), and (4) Slashdot is a social network contains friend/foe links between the users in Slashdot (Leskovec et al., 2008).

These four datasets are relatively large and we would draw samples from these real-world datasets to estimate our target property. Basic information of the real-world datasets is summarized in the Table 1, including the maximum degree of the dataset, so that we can adjust the parameter $C$ according to the maximum degree of the networks. We also give a total variation distance (TVD) (Bar-Yossef and Gurevich, 2008) between the stationary distribution of SRW on the networks and uniform distribution. As we know that the probability of being reached with simple random walker is proportional to the node degree, and any proposed methods are also related to the node degree due to RW basis. Therefore, listing the TVD of them can make a clearer relationship between the dataset themselves and the accuracy of estimation.

In the first experiment, we simulate sampling algorithms on the large-scale network from SNAP and estimate the degree distribution, which is a basic graph property. To assess the accuracy of the estimated degree distribution, we would use the KL-Divergence to assess the errors. We set different budget as 1000, 2000, 3000, 4000, and 5000, which is the nodes that collected by one launch. As the KL-Divergence of 4 different algorithms vary too far, we use the log (KL-Divergence) as the final results. The parameter $C$, as the numerical simulation results shown in (Li et al., 2015) states that GMD performs best when $C$ is around half of maximum degree of the graph, we fix the parameter $C$ as a half of the maximum degree to guarantee that GMD can perform in its best status. In addition, to guarantee the accuracy, we run every algorithm for 100 times and take the average error of all 100 runs. All the results presented in the following contents is the average values.

In the second part, we employ GMD and WJRW on 4 large-scale network datasets, changing the parameter $C$ and get the errors of estimated degree distributions.

*Table 1 Summary of the datasets*

| **Dataset** | **#Nodes** | **#Edges** | $d_{max}$ | **TVD** |
|---|---|---|---|---|
| Wiki-crocodile | 11631 | 170773 | 3546 | 0.473 |
| Slashdot | 70999 | 365572 | 2510 | 0.608 |
| DBLP | 317080 | 1049866 | 343 | 0.400 |
| YouTube | 1134890 | 2987624 | 28754 | 0.578 |

### 4.2 Experimental Results

**Exp 1— Which algorithm produces the most accurate outcome?**

First, we compare the error of different basic random walk samplers which are the Simple Random Walk (SRW), the Random Walk with Escape, and the Generalised Maximum Degree (GMD), and the method proposed in our paper Weighted Jump Random Walk. The results with increasing sample size are reported in Fig. 2. As the plots show, the error of WJRW is constantly better than the other algorithms, which confirm the theoretical analysis presented above. Also, the error of all the algorithms decreases with increasing sample size, which is consistent with the law of large number (LLN). SRW performs poorly, which is consistent with the previous experimental results (Gjoka et al., 2010; Li et al., 2015, 2019; Ribeiro and Towsley, 2010; Zhao et al., 2019).

After obtaining the sample nodes, we also count the number of nodes in the sample after removing the repeated sample nodes. The results are presented in Table 2, which shows that there are less repeated nodes reached by WJRW than SRW, GMD and RWE, thus the weighted jump random walker can explore the graph more exhaustively. This result also implies that WJRW does a better job at overcoming the situation of being stuck in a single community than the other 3 algorithms.

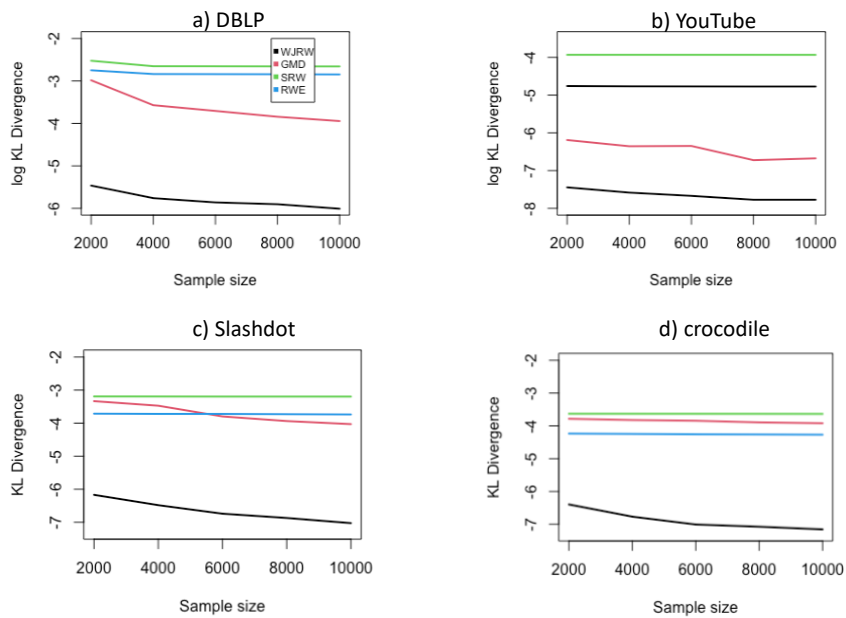

Figure 2 Comparison of Random walk based samplers

Table 2 The number of unique node in sample under specific budget B

|  | B = 2000 | B = 4000 | B = 6000 | B = 8000 | B = 10000 |
|---|---|---|---|---|---|
| GMD | 1204 | 3001 | 4873 | 6648 | 6971 |
| WJRW | 1845 | 3643 | 5291 | 6902 | 8023 |
| SRW | 779 | 1054 | 3018 | 3683 | 4219 |
| RWE | 1194 | 2754 | 3676 | 5459 | 5993 |

**Exp 2— How does parameter $C$ affect the error?**

In the experiment 2, we increase the $C$ from 0 to the $d_{max}$. This setting is the same as the experiment conducted in previous research by Li et al. (2015). They chose this range of C as GMD's stationary distribution becomes uniform and will never change after $C$ exceeding $d_{max}$. There is no need for GMD to increase $C$ to a larger number than the maximum degree of $G$. However, in our method, the stationary distribution is always affected by the degree of nodes, thus it varies with a changing $C$ when $C$ exceeds maximum degree. For practicality, we still use a parameter $C$ smaller than $d_{max}$, so our experiments also adjust $C$ from 0.1 times $d_{max}$ to $d_{max}$.

The results with varying $C$ is in the Fig. 3. Summarizing all curves of real-world datasets in Fig.3, when we use $C$ between 0.4-0.7 times maximum degree, the accuracy is relatively high than other values. For large networks, the range of $C$ is large. For example, Slashdot network has the maximum degree 2510, the suggested choice of $C$ is [1004,1751]. There are over seven hundred choices for the parameter $C$.

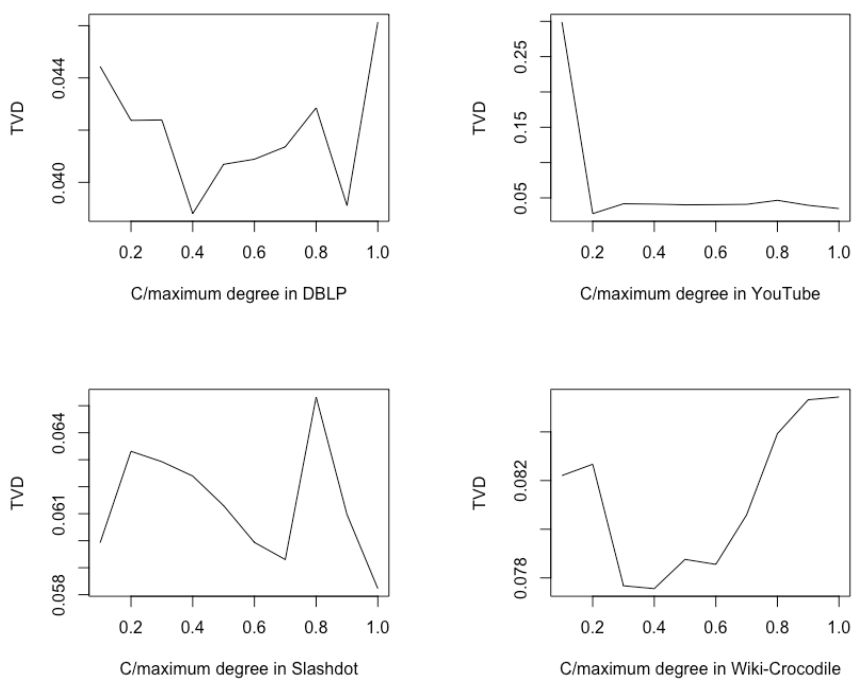

*Figure 3 Estimation accuracy vs. C/maximum degree*

Now we are going to explain the behavior of WJRW after C exceeds $d_{max}$. Once $C$ exceeds $d_{max}$, i.e., all nodes in the graph are in $S$, all the nodes are connected with the virtual node, and the jump probability is deterministic for a known $C$. As the $C$ grows, the stationary probability of node $u$ will be dominated by $\frac{\sum_{u \in V} \alpha_u}{\sum_{u \in V} \mathbb{I}\{d_u < C\}}$, instead of the degree of nodes. In addition, larger $C$ leads to a larger $\alpha$, while the number of nodes in a graph is unchanged. So, the value of

$\frac{\sum_{u \in V} \alpha_u}{\sum_{u \in V} \mathbb{I}\{d_u < C\}}$ will grow larger as $C$ grows larger. When $C$ is approaching infinity, the stationary distribution is approaching uniform. This long-time behavior is similar with RWE but have a more effective improvement in performance. By changing the value of $C$, we can unify the SRW and a uniform vertex sampling into one framework.

### 4.3 Discussion

**Compare with MD/ GMD.** In random walk based graph sampling methods, there are two important principles: 1) correct the large deviation of stationary distribution caused by uniform walk in SRW. 2) collect as much distinct sample nodes as possible. Therefore, many random walk based sampling algorithms aim to increase the probability of low-degree nodes being chosen, either by designing a weighted walk, or increasing the degree of low-degree nodes, so that the newly assigned probabilities is lifted. MD/GMD, WJRW and RWE belong to the latter category.

The initial idea of WJRW starts from overcoming the shortcomings of GMD, alleviating the repeated sample problem caused by self-loop in GMD. Instead of choosing the self-loops on one node, we choose to collect them connected altogether through a virtual node, which seems a small difference. But why GMD still have significantly lower accuracy than WJRW? The answer is WJRW reduced the problem of repeated sample in the very beginning. Repeated problems of MD and GMD lead to low-quality sample. In MD, the estimator takes all sample nodes (including repeated sample) to estimate the target function, but in GMD, the estimator of GMD eliminates the sample nodes collected by consecutive self-loop in the estimator, leading to a small actual sample size, thus the estimation accuracy is restricted.

It is obvious that there is a dilemma of GMD that with small $C$, the stationary distribution is too close to that of SRW, having large deviation problems; with large $C$, although the stationary distribution gets rid of the large deviation, the estimator will eliminate a large proportion of samples (large $C$ means large proportion nodes have self-loops and for one node the specific number of self-loops $C - d_i$ would increase), leading to a waste of sample size. This dilemma constraints the performance of GMD, which is verified in the previous research (Li et al., 2015).

**Compare with RWE/ SRW.** SRW's stationary distribution is proportional to their node degrees. although setting a fixed parameter $\alpha$ can make the stationary distribution smoother, or closer to uniform. However, the mechanism of RWE is to elevate all nodes in the graph, leading to a costly algorithm. Therefore, we target to improve the nodes with low degrees, instead of adding jumping parameter alpha to all nodes with a fixed value, which is what RWE did. Then, our algorithm is less costly than RWE and the stationary distribution can be improved by a smaller node set, which is not expensive to uniformly choose from. Although WJRW is analogous to RWE, which links all nodes in the graphs and setting a fixed probability to jump uniformly. What makes WJRW better than RWE is WJRW targets to elevate the probability of low-degree nodes only, instead of linking high-degree nodes with the virtual node as well. This difference makes WJRW more effective in collecting samples.

Stationary distribution plays an important role in RW-based sampling methods, the closer to uniform it is, the more accurate the estimated value is. Li et al. (2015) give a proof that the similarity between stationary distribution of RW-based methods and uniform distribution determines the effectiveness of them. The smaller the difference is, the more effective the RW-based methods are. Although MD and GMD can get distributions similar to uniform, or even run uniform distributions on target graphs, the existence of self-loop in the algorithms leads to repeated sample problem thus result in low accuracy. The design of self-loop in MD and GMD, is to increase the node degree by linking themselves. However, the self-loop does not only increase the probability of being reached, it also makes the random walker get stuck in the single node, especially when the difference between $C$ and original degree is large. Therefore, linking low-degree nodes with each other can overcome the deficiency of GMD, which obviously has a smaller probability of choosing the same node in 2 consecutive steps. The WJRW essentially links all low-degree nodes by weighted edges, sharing the same weight, which is totally determined by the degree of low-degree nodes. The conductance of the graph is increased through added edges, so that the WJRW can explore the graphs more exhaustively than the other methods.

## 5. Conclusion

In this paper, we analyzed the drawbacks of previous methods, and propose a new method called Weighted Jump Random Walk (WJRW), improving large deviation problem by tying low-degree nodes together, so that the low-degree nodes are more likely to be reached and the conductance of the graph is higher than the original graph. We also conduct extensive experiments with real-world networks, and the result show that our sampling approach diffuses quickly thus samples more exhaustive nodes given a specific budget. Therefore, the accuracy of the estimation based on the sample is higher than various classical sampling algorithms.

# Reference:

Adamic, L.A., Lukose, R.M., Puniyani, A.R., Huberman, B.A., 2008. Search in Power-Law Networks. Phys. Rev. E 64, 046135. https://doi.org/10.1103/PhysRevE.64.046135

Ahn, Y.-Y., Han, S., Kwak, H., Moon, S., Jeong, H., 2007. Analysis of topological characteristics of huge online social networking services, in: Proceedings of the 16th International Conference on World Wide Web - WWW '07. Presented at the the 16th international conference, ACM Press, Banff, Alberta, Canada, p. 835. https://doi.org/10.1145/1242572.1242685

Bar-Yossef, Z., Berg, A., Chien, S., Fakcharoenphol, J., Weitz, D., 2000. Approximating Aggregate Queries about Web Pages via Random Walks, in: Proceedings of the 26th International Conference on Very Large Data Bases, VLDB '00. Morgan Kaufmann Publishers Inc., San Francisco, CA, USA, pp. 535–544.

Bar-Yossef, Z., Gurevich, M., 2008. Random Sampling from a Search Engine's Index 78.

Bianchini, M., Gori, M., Scarselli, F., 2005. Inside PageRank.

Bos, W. van den, Crone, E.A., Meuwese, R., Güroğlu, B., 2018. Social network cohesion in school classes promotes prosocial behavior. PLOS ONE 13, e0194656. https://doi.org/10.1371/journal.pone.0194656

Charitou, T., Bryan, K., Lynn, D.J., 2016. Using biological networks to integrate, visualize and analyze genomics data. Genet Sel Evol 48, 27. https://doi.org/10.1186/s12711-016-0205-1

Duan, Y., Lu, F., 2014. Robustness of city road networks at different granularities. Physica A: Statistical Mechanics and its Applications 411, 21–34. https://doi.org/10.1016/j.physa.2014.05.073

Gjoka, M., Kurant, M., Butts, C.T., Markopoulou, A., 2010. Walking in Facebook: A Case Study of Unbiased Sampling of OSNs, in: 2010 Proceedings IEEE INFOCOM. Presented at the IEEE INFOCOM 2010 - IEEE Conference on Computer Communications, IEEE, San Diego, CA, USA, pp. 1–9. https://doi.org/10.1109/INFCOM.2010.5462078

Gupta, L., Jain, R., Vaszkun, G., 2016. Survey of Important Issues in UAV Communication Networks. IEEE Commun. Surv. Tutorials 18, 1123–1152. https://doi.org/10.1109/COMST.2015.2495297

Horvitz, D.G., Thompson, D.J., 1952. A Generalization of Sampling Without Replacement from a Finite Universe. Journal of the American Statistical Association 47, 663–685. https://doi.org/10.1080/01621459.1952.10483446

Jin, L., Chen, Y., Hui, P., Ding, C., Wang, T., Vasilakos, A.V., Deng, B., Li, X., 2011. Albatross Sampling: Robust and Effective Hybrid Vertex Sampling for Social Graphs.

Leskovec, J., Faloutsos, C., 2006. Sampling from large graphs, in: Proceedings of the 12th ACM SIGKDD International Conference on Knowledge Discovery and Data Mining - KDD '06. Presented at the the 12th ACM SIGKDD international conference, ACM Press, Philadelphia, PA, USA, p. 631. https://doi.org/10.1145/1150402.1150479

Leskovec, J., Kleinberg, J., Faloutsos, C., 2007. Graph evolution: Densification and shrinking diameters. ACM Trans. Knowl. Discov. Data 1, 2. https://doi.org/10.1145/1217299.1217301

Leskovec, J., Lang, K.J., Dasgupta, A., Mahoney, M.W., 2008. Community Structure in Large Networks: Natural Cluster Sizes and the Absence of Large Well-Defined Clusters.

Li, R.-H., Yu, J.X., Qin, L., Mao, R., Jin, T., 2015. On random walk based graph sampling. Presented at the 2015 IEEE 31st International Conference on Data Engineering (ICDE), IEEE Computer Society, pp. 927–938.
https://doi.org/10.1109/ICDE.2015.7113345

Li, Y., Wu, Z., Lin, S., Xie, H., Lv, M., Xu, Y., Lui, J.C.S., 2019. Walking with Perception: Efficient Random Walk Sampling via Common Neighbor Awareness, in: 2019 IEEE 35th International Conference on Data Engineering (ICDE). Presented at the 2019 IEEE 35th International Conference on Data Engineering (ICDE), IEEE, Macao, Macao, pp. 962–973. https://doi.org/10.1109/ICDE.2019.00090

McLaren, C.D., Bruner, M.W., 2022. Citation network analysis. International Review of Sport and Exercise Psychology 15, 179–198. https://doi.org/10.1080/1750984X.2021.1989705

Newman, M.E.J., 2001. The structure of scientific collaboration networks. Proc. Natl. Acad. Sci. U.S.A. 98, 404–409. https://doi.org/10.1073/pnas.98.2.404

Portenoy, J., Hullman, J., West, J.D., 2017. Leveraging Citation Networks to Visualize Scholarly Influence Over Time. Front. Res. Metr. Anal. 2, 8. https://doi.org/10.3389/frma.2017.00008

Ribeiro, B., Towsley, D., 2010. Estimating and Sampling Graphs with Multidimensional Random Walks. arXiv:1002.1751 [cs].

Rozemberczki, B., Allen, C., Sarkar, R., 2021. Multi-Scale attributed node embedding. Journal of Complex Networks 9, cnab014. https://doi.org/10.1093/comnet/cnab014

Rozemberczki, B., Sarkar, R., 2020. Characteristic Functions on Graphs: Birds of a Feather, from Statistical Descriptors to Parametric Models.

Shimbel, A., 1953. Structural parameters of communication networks. Bulletin of Mathematical Biophysics 15, 501–507. https://doi.org/10.1007/BF02476438

Xie, F., Levinson, D., 2007. Measuring the Structure of Road Networks. Geographical Analysis 39, 336–356. https://doi.org/10.1111/j.1538-4632.2007.00707.x

Yang, J., Leskovec, J., 2012. Defining and Evaluating Network Communities based on Ground-truth.

Zhang, P., Itan, Y., 2019. Biological Network Approaches and Applications in Rare Disease Studies. Genes 10, 797. https://doi.org/10.3390/genes10100797

Zhao, J., Wang, P., Lui, J.C.S., Towsley, D., Guan, X., 2019. Sampling online social networks by random walk with indirect jumps. Data Min Knowl Disc 33, 24–57. https://doi.org/10.1007/s10618-018-0587-5